\documentclass[reprint,amsmath,amssymb,prl,superscriptaddress]{revtex4-2}

\usepackage{dcolumn}
\usepackage{bm}
\usepackage{dsfont}
\usepackage{graphicx,epsfig}
\usepackage{subfigure}
\usepackage{amsmath}
\usepackage{braket}
\usepackage{caption}
\usepackage{float}
\usepackage[shortlabels]{enumitem}
\usepackage{hyperref}
\usepackage{blindtext}
\usepackage{color}

\DeclareMathOperator{\Tr}{Tr}

\renewcommand{\eqref}[1]{(\ref{#1})}
\newcommand{\figref}[1]{Fig.~\ref{#1}}
\newcommand{\appref}[1]{App.~\ref{#1}}

\begin{document}

\title{Quantum vs.~noncontextual semi-device-independent randomness certification}

\author{Carles Roch i Carceller}\email{crica@dtu.dk}\address{Department of Physics, Technical University of Denmark, Fysikvej, 2800 Kgs. Lyngby, Denmark}

\author{Kieran Flatt}\affiliation{School of Electrical Engineering, Korea Advanced Institute of Science and Technology (KAIST),291 Daehak-ro Yuseong-gu, Daejeon 34141 Republic of Korea}

\author{Hanwool Lee}\affiliation{School of Electrical Engineering, Korea Advanced Institute of Science and Technology (KAIST),291 Daehak-ro Yuseong-gu, Daejeon 34141 Republic of Korea}

\author{Joonwoo Bae}\affiliation{School of Electrical Engineering, Korea Advanced Institute of Science and Technology (KAIST),291 Daehak-ro Yuseong-gu, Daejeon 34141 Republic of Korea}

\author{Jonatan Bohr Brask}\address{Department of Physics, Technical University of Denmark, Fysikvej, 2800 Kgs. Lyngby, Denmark}

\begin{abstract}
We compare the power of quantum and classical physics in terms of randomness certification from devices which are only partially characterised. We study randomness certification based on state discrimination and take noncontextuality as the notion of classicality. A contextual advantage was recently shown to exist for state discrimination. Here, we develop quantum and noncontextual semi-device independent protocols for random-number generation based on maximum-confidence discrimination, which generalises unambiguous and minimum-error state discrimination. We show that, for quantum eavesdropppers, quantum devices can certify more randomness than noncontextual ones whenever none of the input states are unambiguously identified. That is, a quantum-over-classical advantage exists.
\end{abstract}

\maketitle


Quantum physics departs radically from everyday experience. Observations on quantum systems can defy classical notions of cause and effect and exploiting quantum effects enables advantages for a number of applications including precision sensing, computing, and information security. Understanding the quantum-classical boundary is both of fundamental importance to the foundations of physics in general and of relevance to characterising and quantifying quantum-over-classical advantages in specific tasks and applications.

In this work, we compare the power of quantum and classical physics for randomness certification. Random numbers are needed for many tasks in science and technology \cite{Hayes2001,Bera2017}. In particular, high-quality randomness is central to cryptographic security and thus to much of modern information technology. Due to the inherent randomness in quantum measurements, strong guarantees can be established for the extraction of randomness from quantum systems. In fact, randomness can be certified with little or no trust in the devices used to generate it. In setups with multiple, separate parties, randomness can be certified in a device-independent (DI) setting, where the devices are treated as untrusted black boxes \cite{colbeckPhD2009,Pironio2010,Acin2016}. In that setting, the relevant notion of classicality is locality (also known as local causality), in the sense of Bell \cite{Bell1964,Brunner2014}, and the setup is required to violate a Bell inequality to generate randomness. This is, however, technologically very demanding, as the violation must be loophole free \cite{Pironio2010,Christensen2013,Liu2018,Bierhorst2018,Shalm2021,Liu2021}. Here, we focus on the semi-DI setting, where the black boxes are complemented by a few, general assumptions, representing an increased level of trust in the devices. This renders implementations much more accessible, and semi-DI randomness certification can be realised in simple prepare-and-measure setups \cite{Li2011,Vallone2014,Lunghi2015,Mironowicz2021,Cao2015,Marangon2017,Cao2016,Xu2016,Brask2017a,Michel2019,Rusca2019,Drahi2020}. As our notion of classicality we adopt noncontextuality \cite{Kochen1968,Budroni2021}, in the form introduced by Spekkens \cite{Spekkens2005}, which is applicable also in scenarios which do not have the multipartite structure of Bell tests.

We consider semi-DI randomness certification based on state discrimination, where the partial trust in the devices consists in an assumption about the distinguishability of the prepared states. In particular, we consider maximum-confidence state discrimination \cite{Croke2006}. In the context of randomness certification, a semi-DI protocol based on unambiguous state discrimination was previously demonstrated \cite{Brask2017}, and in the context of comparing quantum and noncontextual models, a quantum advantage for minimum-error state discrimination was demonstrated by Schmid and Spekkens \cite{Schmid2018}. Maximum-confidence discrimination is more general, containing minimum-error and unambiguous state discrimination as particular cases. In related work, we demonstrate a quantum-over-noncontextual advantage for maximum-confidence state discrimination \cite{Flatt2021}. In the present work, we find a rich picture. In a setting where the devices are either quantum or noncontextual, but where the eavesdropper in both cases is allowed quantum powers, quantum devices outperform noncontextual ones. However, comparing a quantum universe with quantum eavesdroppers against a noncontextual universe with noncontextual (hence less powerful) eavesdroppers, the amount of quantum certifiable randomness may be both larger than, smaller than or equal to the amount of noncontextual randomness, depending on the distinguishability of the states and the observed confidence of discrimination.

\begin{figure*}
    \centering
    \includegraphics[width=\textwidth]{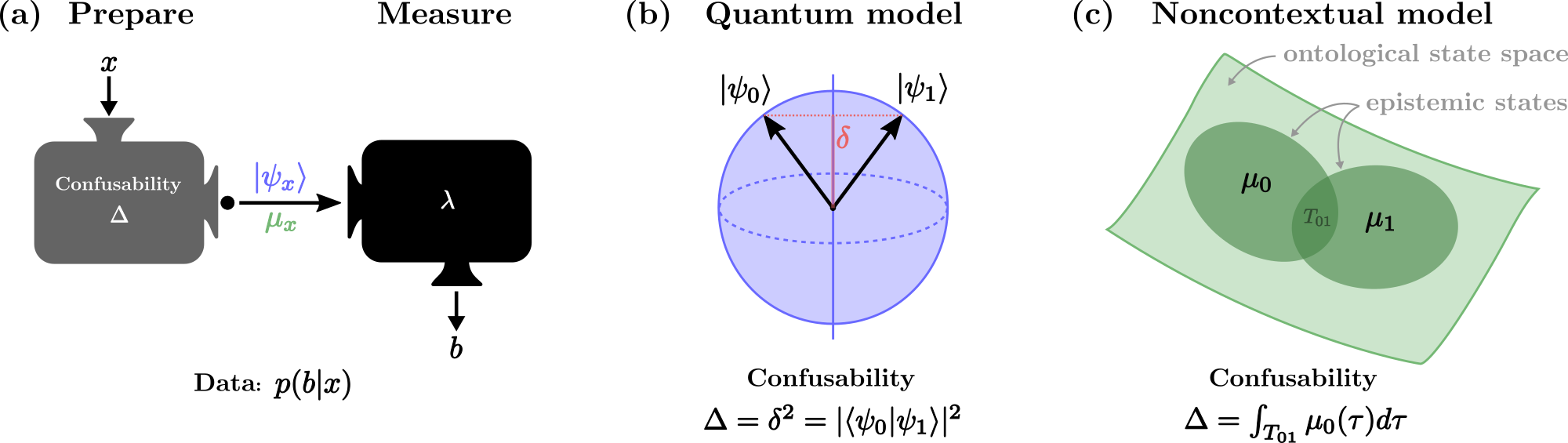}
    \caption{\textbf{(a)} Prepare-and-measure scenario for state discrimination and randomness certification, in quantum and non-contextual settings. A preparation device takes an input and transmits states to a measurement device, which produces an output. From an assumption about the distinguishability of the states and the observed input-output correlations, the entropy in the raw output can be bounded and random numbers extracted from it. \textbf{(b)} In the quantum setting, the distinguishability is quantified by the overlap of the quantum states. For binary inputs, these can be represented by qubit states. \textbf{(c)} In the non-contextual setting, there is an ontological state space, consisting of perfectly distinguishable states. The preparation device emits epistemic states, given by probability distributions over ontological states. The distinguishability of epistemic states is quantified by the confusability, which measures the overlap of the corresponding distributions. }
    \label{fig.concept}
\end{figure*}


A prepare-and-measure setting for state discrimination and randomness certification is illustrated in \figref{fig.concept}(a). We will restrict our attention to binary inputs $x \in \{0,1\}$ and ternary outputs $b \in \{0,1,\o\}$. In the case of state discrimination, $b$ represents a guess for which state was prepared, with $\o$ labelling inconclusive outcomes. For randomness certification, the amount of true randomness present in the output $b$ can be lower bounded based on the observed distribution $p(b|x)$ and an assumption on the distinguishability of the prepared states. We start by considering state discrimination, first in the quantum case and then for noncontextual theories.

In quantum state discrimination, quantum states $\rho_x$ are prepared and the measurement device implements a POVM with elements $\Pi_b$, resulting in the distribution $p(b|x) = \Tr[\rho_x \Pi_b]$. For binary inputs, without loss of generality, the state space can be taken to be a qubit space. When the states are furthermore pure, $\rho_x = \ket{\psi_x}\bra{\psi_x}$, their distinguishability can be quantified simply by their overlap $\delta = |\braket{\psi_0|\psi_1}|$. Different quantifiers of performance can be adopted.

In minimum-error state discrimination (MESD), no inconclusive outcomes are permitted, $p(\o|x) = 0$, and the figure of merit is the average error rate $P_{e} = p_0 p(1|0) + p_1 p(0|1)$, where $p_x$ is the prior probability for input $x$. Optimal MESD achieves a minimal error rate given by the Helstrom bound $P_e = \frac{1}{2}(1 - \sqrt{1-4 p_0 p_1 \delta^2})$ \cite{Helstrom1976}. Thus, errors are unavoidable for non-orthogonal states.

Errors can be suppressed at the cost of a non-zero rate of inconclusive outcomes. In unambiguous state discrimination (USD), the error probabilities are strictly zero, $p(0|1)=p(1|0)=0$, and the average inconclusive rate $P_{\o} = p_0 p(\o|0) + p_1 p(\o|1)$ can be taken as the figure of merit. For unbiased inputs, $p_0=p_1=\frac{1}{2}$, optimal USD achieves $P_{\o} = \delta$ \cite{Barnett2009}. In the case of qubits, USD is possible only for two pure states. 

Maximum-confidence discrimination (MCD) generalises the notions of MESD and USD \cite{Croke2006}. The \textit{confidence} $C_x$ is the probability that, given an outcome $b=x$, the input was $x$. From Bayes' theorem
\begin{equation}
    \label{eq.confdef}
    C_x = \frac{p_x}{\eta_x} p(x|x) ,
\end{equation}
where $\eta_b = \sum_x p(b|x) p_x$ is the rate of outcome $b$ (i.e.~the marginal distribution of the output). In MCD, the figure of merit is a given $C_x$, or any convex combination of them, and the goal is to maximise this quantity. When $C_x=1$, the input $x$ is unambiguously identified. Hence, unambiguous discrimination is a particular case of MCD, and if no further constraints are imposed, MCD recovers USD whenever the latter is possible. This is the case for an arbitrary number of linearly independent pure states, and thus in particular always for two distinct pure states, as considered here. MESD can also be recovered by adopting $\eta_0 C_0 + \eta_1 C_1 = 1 - P_e$ as the figure of merit, when the inconclusive rates are zero \cite{Barnett2009}. In general, MCD is flexible and can handle situations in which both error rates and inconclusive rates are nonzero.

\begin{figure*}
    \includegraphics[width=1.1\textwidth,trim={2cm 0.5cm 0 0},clip]{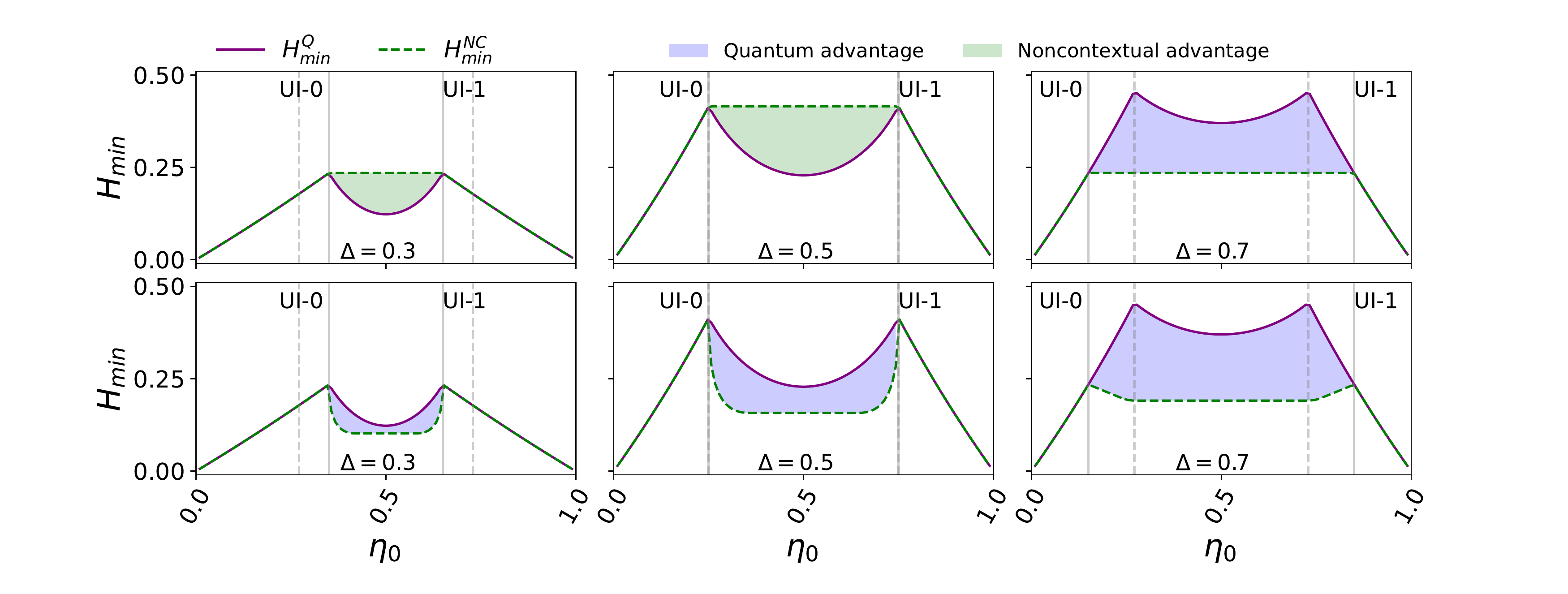}
    \caption{Quantum $H^Q_{min}$ and noncontextual $H^{NC}_{min}$ certifiable min-entropies vs.~output rate $\eta_0$, for three different confusabilities $\Delta$, optimal confidence $C_0$, and equal prior probabilities $p_0=p_1=\frac{1}{2}$. Solid vertical lines delimit parameter regions in which input $x$ is unambiguously identified, labelled UI-$x$. Dashed vertical lines indicate rates at which $H_{min}^{Q}$ is maximal. The confidences are maximal in all plots. Top row: eavesdroppers in quantum and noncontextual models are respectively quantum and noncontextual. Bottom row: a quantum eavesdropper is considered in both cases.}
    \label{fig.Hmincompare}
\end{figure*}

We now proceed to consider noncontextual state discrimination. We start from an ontological model of the prepare-and-measure scenario \cite{Spekkens2008,Schmid2018}. The system is associated with an ontic state space $T$ in which each point $\tau$ completely defines all physical properties, i.e.~the outcomes of all possible measurements. Each state preparation $x$ samples the ontic state space according to a probability distribution $\mu_{x} (\tau)$, referred to as the epistemic state. Each measurement is defined by a set of response functions, that is, non-negative functions $\xi_b(\tau)$ over the ontic space, such that $\sum_b \xi_b(\tau) = 1$ for all $\tau \in T$. The probability of obtaining the outcome $b$ when state $\mu_x$ was prepared is
\begin{equation}
\label{eq.ontprob}
p(b|x) = \int_T d\tau \mu_x(\tau) \xi_b(\tau) .
\end{equation}
While distinct ontic states can be perfectly discriminated, epistemic states with overlapping distributions cannot. It is the discrimination of epistemic states which we compare against quantum state discrimination. 

To compare the two requires a notion analogous to the quantum state overlap. Note that $\delta^2 = |\braket{\psi_0|\psi_1}|^2$ can be thought of as the probability that an outcome corresponding to projection onto $\ket{\psi_1}$ occurs when $\ket{\psi_0}$ was prepared (or vice versa). Similarly, in the ontological model we define sharp outcomes as outcomes that are certain to occur for a given preparation. $\xi_b$ is a sharp outcome for $\mu_x$ if $p(b|x) = 1$. For discrimination of $\mu_0$ and $\mu_1$, the \textit{confusability} $\Delta_{0,1}$ is then the probability that a sharp outcome for $\mu_1$ occurs when $\mu_0$ was prepared. For preparation-noncontextual models, that we now introduce, one has the same symmetry as in the quantum case $\Delta_{0,1} = \Delta_{1,0} = \Delta$, and the models can be compared for $\Delta=\delta^2$. 

Two preparation procedures are said to be operationally equivalent if they cannot be distinguished by any measurement, and the ontological model is said to be \textit{preparation noncontextual} if all operationally equivalent preparations are represented by the same epistemic state. We take preparation noncontextuality as our notion of classicality and refer to it simply as noncontextuality. We impose two requirements on the noncontextual model. First, it reproduces the observed distribution $p(b|x)$. Second, we need an operational equivalence to which noncontextuality can be applied. We take the mode to reproduce the existence of complementary states $\ket{\psi_{\bar{x}}}$, with $\ket{\psi_x}\bra{\psi_{x}} + \ket{\psi_{\bar{x}}}\bra{\psi_{\bar{x}}} = \openone$ and $|\braket{\psi_{\bar{0}}|\psi_{\bar{1}}}| = \delta$. That is, in addition to the epistemic states $\mu_0$, $\mu_1$, it must also contain two states $\mu_{\bar{0}}$, $\mu_{\bar{1}}$ such that their confusability is $\Delta$, they obey $\mu_x \mu_{\bar{x}} = 0$, and the convex combinations $\frac{1}{2}\mu_x + \frac{1}{2}\mu_{\bar{x}}$ for $x=0,1$ correspond to operationally equivalent preparations. By noncontextuality they must hence be equal $\frac{1}{2}\mu_0 + \frac{1}{2}\mu_{\bar{0}} = \frac{1}{2}\mu_1 + \frac{1}{2}\mu_{\bar{1}}$. It was shown by Schmid and Spekkens, under similar assumptions, that quantum mechanics outperforms noncontextual theory for MESD in the sense that the Helstrom bound is lower than the minimum achievable error rate in the noncontextual model for any value of $\Delta$ \cite{Schmid2018}. In Ref.~\cite{Flatt2021}, we study quantum vs.~noncontextual maximum-confidence discrimination.

The prepare-and-measure state-discrimination setup can be exploited for semi-DI randomness certification by taking $\Delta$ as given while the devices are otherwise uncharacterised (the states and measurements are unknown), and then assess the randomness of $b$ based on the observed distribution $p(b|x)$. Intuitively, if $p(b|x)$ is close to optimal discrimination for the given $\Delta$, this constrains the measurements to be close to the optimal ones, and the predictability of $b$ to someone with perfect knowledge of the states and measurements can be estimated. More precisely, we introduce a hidden variable $\lambda$, distributed according to $q_\lambda$, labelling measurement strategies. The average guessing probability for an eavesdropper with access to $\lambda$ and the input $x$
\begin{equation}
\label{eq:pg_q}
    p_g = \sum_x p_x \sum_\lambda q_{\lambda} \max_b p(b|x,\lambda) ,
\end{equation}
with $p(b|x,\lambda)$ given by $\Tr[\rho_x \Pi^\lambda_b]$ when the eavesdropper is quantum and by \eqref{eq.ontprob} with response function $\xi^\lambda_b$ if the eavesdropper is restricted to be noncontextual. Note that $\lambda$ is assumed to be independent of $x$ (otherwise the discrimination problem becomes trivial). We quantify the randomness by the min-entropy $H_{min} = -\log_2 p_g$, which gives the number of (almost) uniformly random bits which can be extracted per round of the protocol \cite{Konig2009}.

Since the measurement strategies are unknown to the user, to certify randomness $p_g$ must be upper-bounded by optimising over all strategies compatible with the observed data. We focus on MCD for the input $x=0$ and impose only that the rate $\eta_0$ and the confidence $C_0$ are reproduced (as opposed to the full distribution $p(b|x)$. For a quantum eavesdropper, $p_g \leq p_g^Q$ with
\begin{equation}
\label{eq.Qopt}
    p_g^Q = \max_{q_{\lambda},\Pi^\lambda_b} \sum_{x,\lambda} p_x q_{\lambda} \max_b \Tr[\hat{\rho}_x \hat{\Pi}^\lambda_b] ,
\end{equation}
subject to $q_\lambda$ and $\Pi^\lambda_b$ being valid probability distributions and POVMs respectively, $\sum_{x,\lambda} q_{\lambda} p_x \Tr[\hat{\rho}_x \hat{\Pi}^{\lambda}_0] = \eta_0$ and $\sum_{\lambda} q_{\lambda} p_0 \Tr[\hat{\rho}_0 \hat{\Pi}^{\lambda}_0] = \eta_0 C_0$. Without loss of generality, the states can be fixed to any pair of states with overlap $\delta$. Thus $p_g^Q$ is a function only of the confusability $\Delta$ and the distribution $p(b|x)$. The optimisation problem in \eqref{eq.Qopt} can be rendered as a semidefinite program, as we show in \appref{ap:QRNG}.

Similarly, the guessing probability for a noncontextual eavesdropper is bounded by $p_g \leq p_g^{NC}$ with
\begin{equation}
\label{eq.NCopt}
    p_g^{NC} = \max_{q_{\lambda},M^\lambda_b} \sum_{x,\lambda} p_x q_{\lambda} \max_b \int_T d\tau \mu_x(\tau) \xi^\lambda_b(\tau) ,
\end{equation}
where now $\xi^\lambda_b$ must be valid response functions, and the constraints are the same as in the quantum case with the Born rule replaced by \eqref{eq.ontprob}.

In a noncontextual theory, a pair of epistemic states must be equal on the overlap of their supports \cite{Spekkens2005,Schmid2018}. This allows a general response function to be decomposed into four extremal functions, corresponding to integrals over the regions defined by the overlapping supports of $\mu_0$, $\mu_1$ and their non-overlapping partners. These integrals are, furthermore, functions of the confusability $\Delta$. Using this, in \appref{ap:RNG_NC} we show that \eqref{eq.NCopt} can also be rendered as a semidefinite program.

In \figref{fig.Hmincompare}, we compare the certifiable quantum and noncontextual min-entropies, $H^Q_{min}$ and $H^{NC}_{min}$, in two different manners, focusing on equal prior probabilities $p_0=p_1=\frac{1}{2}$ for simplicity. First, we compute the certifiable $H_{min}$ within each theory (top row), i.e.,~$H_{min}^{Q}$ when the device attains the maximum quantum confidence and the eavesdropper is also quantum, and $H_{min}^{NC}$ for maximum noncontextual confidence and a noncontextual eavesdropper. This is the maximal certifiable randomness in each theory, as $H_{min}$ is maximised for optimal discrimination. Second, we consider the case in which the eavesdropper is always quantum (bottom row). That is, the minimum entropy is computed via the quantum SDP. Since quantum MCD can reach higher confidences than noncontextual MCD, $C_0$ is not necessarily the same in the two cases.

\begin{figure}
    \centering
    \includegraphics[width=0.5\textwidth]{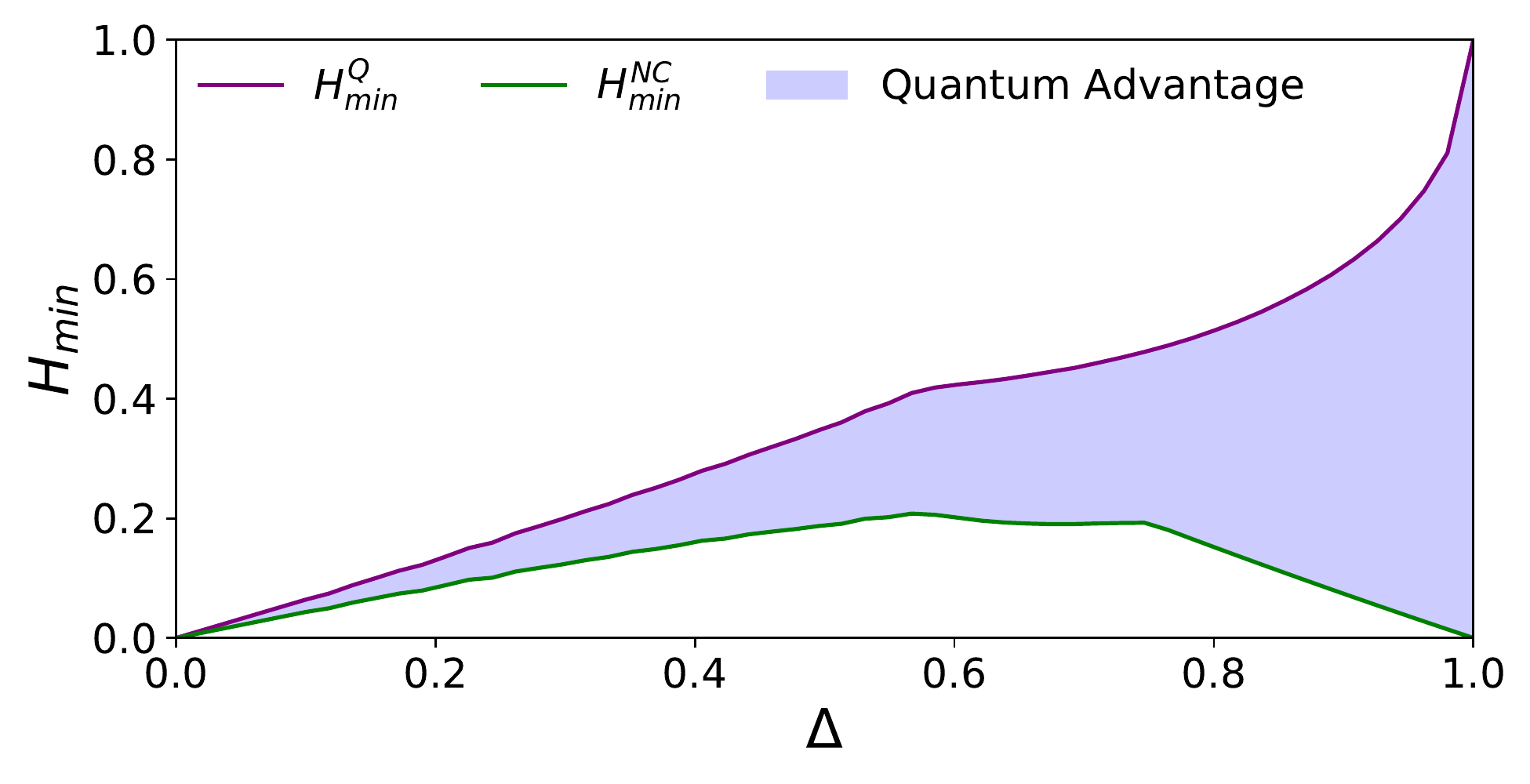}
    \caption{Minimum entropy corresponding to the output rates with maximal quantum advantage, for quantum and noncontextual discrimination schemes and a quantum eavesdropper.}
    \label{fig.maxrand}
\end{figure}

In the first case we find quantum-over-noncontextual as well as noncontextual-over-quantum advantages in terms of certifying randomness. Whenever any of the states is unambiguously identified by the measurement device, the quantum and noncontextual certifiable randomness are equal, $H_{min}^{Q}=H_{min}^{NC}$. Outside these regions, for confusabilities $\Delta < 1/2$ there is a noncontextual advantage, while for $\Delta > 1/2$ a quantum advantage appears and eventually dominates for large $\Delta$. We interpret this as follows. A quantum eavesdropper is more powerful than a noncontextual one, but optimal quantum discrimination also imposes stronger constraints on the measurement device. For states that are easy to discriminate (low $\Delta$), the former effect wins while for states that a hard to distinguish (high $\Delta$), the second effect dominates. Note that a noncontextual advantage appears only in a universe where the eavesdropper is noncontextual, but does not have access to the ontic state.

In the second case, the eavesdropper is quantum in both models, i.e., we allow the eavesdropper in the noncontextual setting more power. As may be expected, quantum devices are then always at least as powerful as noncontextual ones, with a quantum-over-noncontextual advantage appearing for all values of $\Delta$ whenever none of the inputs are unambigiously identified.

The maximal quantum advantage in terms of generating unpredictable (random) measurement outputs for a quantum eavesdropper is plotted against the confusability in \figref{fig.maxrand}. The quantum advantage is largest for nearly indistinguishable states (similar to what was found in Ref.~\cite{Ioannou2019}).

In conclusion, we have computed the amount of randomness which can be semi-device-independently certified in maximum-confidence state discrimination setups in both quantum and preparation-noncontextual models. We have derived the maximal randomness within each model, and we find a quantum advantage for MSD-based randomness generation against quantum adversaries. When the adversary in the noncontextual setting is constrained to be noncontextual as well, we find a quantum advantage when the prepared states are difficult to distinguish, but a noncontextual advantage when they are easy to distinguish. In the future, it would be interesting to extend these results to settings with more than two inputs, where more randomness can potentially be generated, and to mixed-state preparations, where correlations between the prepared states and the eavesdropper potentially need to be taken into account.

\textit{Acknowledgements.} JBB and CRC were supported by the Independent Research Fund Denmark and a KAIST-DTU Alliance stipend. KF, HL, and JB were supported by National Research Foundation of Korea (NRF-2021R1A2C2006309), Institute of Information \& communications Technology Planning \& Evaluation (IITP) grant (Grant No. 2019-0-00831, the ITRC Program/IITP-2021-2018-0-01402).


\bibliography{apssamp}

\clearpage

\appendix

\begin{widetext}

\section{SDP derivation for quantum randomness certification}
\label{ap:QRNG}

In this appendix, we show how that the average guessing probability for a quantum eavesdropper (in \eqref{eq.Qopt}) can be rendered as a semidefinite program (SDP). Our derivation closely follows \cite{Brask2017}. \\

At first glance, the objective function in \eqref{eq.Qopt} is nonlinear in the variables $q_\lambda$, $\hat{\Pi}_{b}^{\lambda}$, contains a maximisation over $b$, and the number of strategies $\lambda$ is a priory unbounded. The latter two issues can be resolved, following \cite{Bancal2014}, by noting that all strategies for which the max occurs for the same $b$ for given $x$ can be lumped together. Hence, only $|b|^|x| = 3^{2}=9$ strategies are required. We label each strategies by $(\lambda_0,\lambda_1)$ where $\lambda_{x}\in\left\{0,1,ø\right\}$ indicates the optimal $b$ given $x$. Thus
\begin{equation}
p_g^Q = \max_{q_{\lambda_0,\lambda_1},\Pi^{\lambda_0,\lambda_1}_b} \sum_{x,\lambda} p_x q_{\lambda} \Tr[\hat{\rho}_x \hat{\Pi}^{\lambda_0,\lambda_1}_{\lambda_x}]    ,
\end{equation}
where the distribution over strategies and the POVM elements fulfill
\begin{align}
\sum_{\lambda_{0},\lambda_{1}}q_{\lambda_{0}\lambda_{1}} &=1, \\
\quad q_{\lambda_{0}\lambda_{1}}\ & \geq 0 \,\, \forall \lambda_{0},\lambda_{1} , \\
\hat{\Pi}_b^{\lambda_{0}\lambda_{1}} & = \left(\hat{\Pi}_{b}^{\lambda_{0}\lambda_{1}}\right)^{\dagger} \,\, \forall \lambda_{0},\lambda_{1}  , \\ \hat{\Pi}_{b}^{\lambda_{0}\lambda_{1}} & \geq 0 \,\, \forall \lambda_{0},\lambda_{1},b , \\
\sum_{b}\hat{\Pi}_{b}^{\lambda_{0}\lambda_{1}} & =\mathds{1} \,\, \forall \lambda_{0},\lambda_{1}  ,
\end{align}
and the observed output rate $\eta_{0}$ and confidence $C_{0}^{Q}$ should be reproduced
\begin{align}
\sum_{\lambda_{0},\lambda_{1}}\sum_{x}p_{x}q_{\lambda_{0}\lambda_{1}}\mathrm{Tr}\left[\hat{\Pi}^{\lambda_{0}\lambda_{1}}_{0}\hat{\rho}_{x}\right] =&\eta_{0} \label{eq:a1d}\\
\sum_{\lambda_{0},\lambda_{1}}\frac{p_{0}}{\eta_{0}}q_{\lambda_{0}\lambda_{1}}\mathrm{Tr}\left[\hat{\Pi}^{\lambda_{0}\lambda_{1}}_{0}\hat{\rho}_{0}\right] =& C_{0}^{Q} \label{eq:a1f} \ .
\end{align}
Next, $p_{g}^{Q}$ and the constraints can be linearised by defining new optimisation variables $\hat{M}_{b}^{\lambda_{0}\lambda_{1}}=q_{\lambda_{0}\lambda_{1}}\hat{\Pi}_{b}^{\lambda_{0}\lambda_{1}}$. The primal version of the SDP can then be written:
\begin{equation}
    \label{eq:primQ}
    \boxed{
    \begin{array}{l}
    \begin{array}{ll}
      \underset{\hat{M}_{b}^{\lambda_{0}\lambda_{1}}}{\mathrm{maximise}}  & \displaystyle p_{g}^{Q} = \sum_{x=0}^{1}\sum_{\lambda_{0},\lambda_{1}}p_{x}\mathrm{Tr}\left[\hat{M}^{\lambda_{0}\lambda_{1}}_{\lambda_{x}}\hat{\rho}_{x}\right]  \\ \\
      \mathrm{subject \ to:}  & 
     \end{array} \\ 
     \begin{array}{l}
       \displaystyle\hat{M}_{b}^{\lambda_{0}\lambda_{1}}\geq 0, \ \left(\hat{M}_{b}^{\lambda_{0}\lambda_{1}}\right)^{\dagger}=\hat{M}_{b}^{\lambda_{0}\lambda_{1}} , \ \forall \lambda_{0},\lambda_{1},b \\ \\
       \displaystyle\sum_{b}\hat{M}_{b}^{\lambda_{0}\lambda_{1}} = \frac{1}{2}\mathrm{Tr}\left[\sum_{b}\hat{M}_{b}^{\lambda_{0}\lambda_{1}}\right]\mathds{1}, \ \forall \lambda_{0},\lambda_{1} \\ \\
      \displaystyle  \sum_{b}\sum_{\lambda_{0},\lambda_{1}}\sum_{x}p_{x}\mathrm{Tr}\left[\hat{M}^{\lambda_{0}\lambda_{1}}_{b}\hat{\rho}_{x}\right] = 1 \\ \\
      \displaystyle\sum_{\lambda_{0},\lambda_{1}}\sum_{x}p_{x}\mathrm{Tr}\left[\hat{M}^{\lambda_{0}\lambda_{1}}_{0}\hat{\rho}_{x}\right] = \eta_{0}  \\ \\
      \displaystyle\sum_{\lambda_{0},\lambda_{1}}\frac{p_{0}}{\eta_{0}}\mathrm{Tr}\left[\hat{M}^{\lambda_{0}\lambda_{1}}_{0}\hat{\rho}_{0}\right] = C_{0}^{Q} \ .
    \end{array}
    \end{array}}
\end{equation}
The last two constraints can be reduced to
\begin{align}
\label{eq:comb}
    \sum_{\lambda_{0},\lambda_{1}}p_{x}\mathrm{Tr}\left[\hat{M}^{\lambda_{0}\lambda_{1}}_{0}\hat{\rho}_{x}\right]
    =  \ \eta_{0}C_{0}^{Q}\delta_{x,0} + \eta_{0}\left(1-C_{0}^{Q}\right)\delta_{x,1} \nonumber \ ,
\end{align}
and normalisation implies
\begin{equation}
    \sum_{b}\sum_{\lambda_{0},\lambda_{1}}\sum_{x}\mathrm{Tr}\left[\hat{M}^{\lambda_{0}\lambda_{1}}_{b}\hat{\rho}_{x}\right] = 2 \ .
\end{equation}
Further on, we formulate the dual version of the problem. From each primal constraint in \eqref{eq:primQ}, we introduce the dual variables $\hat{G}_{b}^{\lambda_{0}\lambda{1}}$, $\hat{H}^{\lambda_{0}\lambda{1}}$, $\nu_{x}$ and $\chi$. The corresponding Lagrangian will then be
\begin{align}
&\mathcal{L} = \sum_{x}\sum_{\lambda_{0},\lambda_{1}}p_{x}\mathrm{Tr}\left[\hat{\rho}_{x}\hat{M}_{\lambda_{x}}^{\lambda_{0}\lambda_{1}}\right]
+\sum_{b}\sum_{\lambda_{0},\lambda_{1}}\mathrm{Tr}\left[\hat{G}_{b}^{\lambda_{0}\lambda_{1}}\hat{M}_{b}^{\lambda_{0}\lambda_{1}}\right] 
+\sum_{\lambda_{0},\lambda_{1}}\mathrm{Tr}\left[\hat{H}^{\lambda_{0}\lambda_{1}}\sum_{b}\left(\hat{M}_{b}^{\lambda_{0}\lambda_{1}}-\frac{1}{2}\mathrm{Tr}\left[\hat{M}_{b}^{\lambda_{0}\lambda_{1}}\right]\mathds{1}\right)\right] \label{eq:lagQ} \\
&+\sum_{x}\nu_{x}\left(\sum_{\lambda_{0},\lambda_{1}}p_{x}\mathrm{Tr}\left[\hat{\rho}_{x}\hat{M}_{0}^{\lambda_{0}\lambda_{1}}\right]-\eta_{0}\left(\delta_{x,0}C_{0}^{Q}+\delta_{x,1}\left(1-C_{0}^{Q}\right)\right)\right)
+\chi\left(\sum_{b}\sum_{\lambda_{0},\lambda_{1}}\sum_{x}p_{x}\mathrm{Tr}\left[\hat{\rho}_{x}\hat{M}_{b}^{\lambda_{0}\lambda_{1}}\right]-1\right) \  \nonumber.
\end{align}
Let us now introduce the supremum of the Lagrangian,
\begin{equation}
    \label{eq:supQ}
    \mathcal{S}\equiv\underset{\hat{M}_{b}^{\lambda_{0}\lambda_{1}}}{\mathrm{supp}}  \mathcal{L} \ .
\end{equation}
Given any solution $\hat{M}_{b}^{\lambda_{0}\lambda_{1}}$ of the primal, the last three terms in \eqref{eq:lagQ} vanish. Thus, as $\hat{M}_{b}^{\lambda_{0}\lambda_{1}}$ are constrained to be positive semi-definite, the first line in \eqref{eq:lagQ} yields an upper bound on the guessing probability $p_{g}^{Q}$ (only if all $\hat{G}_{b}^{\lambda_{0}\lambda_{1}}$ are positive semi-definite). The dual can then be formulated by minimising the supremum in \eqref{eq:supQ}. We re-write it as follows:
\begin{align}
\label{eq:sQr}
    \displaystyle\mathcal{S}=\underset{\hat{M}_{b}^{\lambda_{0}\lambda_{1}}}{\mathrm{supp}} \sum_{\lambda_{0},\lambda_{1}}\mathrm{Tr}\left[\hat{M}_{b}^{\lambda_{0}\lambda_{1}}\hat{K}^{\lambda_{0}\lambda_{1}}_{b}\right]-\sum_{x}\nu_{x}\eta_{0}\left(\delta_{x,0}C_{0}^{Q}+\delta_{x,1}\left(1-C_{0}^{Q}\right)\right)-\chi,
\end{align}
where,
\begin{align}
\displaystyle\hat{K}_{b}^{\lambda_{0}\lambda_{1}}=\sum_{x}p_{x}\hat{\rho}_{x}\left(\delta_{b,\lambda_{x}}+\nu_{x}\delta_{b,0}+\chi\right) +\hat{G}_{b}^{\lambda_{0}\lambda_{1}}+\hat{H}^{\lambda_{0}\lambda_{1}}-\frac{1}{2}\mathrm{Tr}\left[\hat{H}^{\lambda_{0}\lambda_{1}}\right]\mathds{1} \ .
\end{align}
Now the supremum in \eqref{eq:sQr} will diverge, unless  $\hat{K}_{b}^{\lambda_{0}\lambda_{1}}=0$. We will drop $\hat{G}_{b}^{\lambda_{0}\lambda_{1}}$, imposing that the remaining expression is negative. This way, the guessing probability can be upper bounded by:
\begin{equation}
    p_{g}\leq p_{g}^{Q}= -\sum_{x=0}^{1}\nu_{x}\left(\delta_{x,0}C_{0}^{Q}+\delta_{x,1}\left(1-C_{0}^{Q}\right) \right)-\chi
\end{equation}
for a given value of confidence $C_{0}$ in discriminating $\hat{\rho}_{1}$ and any numbers $\nu_{x}$ and $\chi$ which fulfil that there exists four $2\times 2$ hermitian matrices $\hat{H}^{\lambda_{0}\lambda_{1}}$, with indices $\lambda_{0},\lambda_{1}=0,1,ø$, such that:
\begin{align}
\displaystyle\sum_{x=0}^{1}p_{x}\hat{\rho}_{x}\left(\delta_{b,\lambda_{x}}+\nu_{x}\delta_{b,0}+\chi
\right) + \hat{H}^{\lambda_{0}\lambda_{1}}-\frac{1}{2}\mathrm{Tr}\left[\hat{H}^{\lambda_{0}\lambda_{1}}\right]\mathds{1} \leq 0 \ .
\end{align}

\section{Matrix notation for noncontextual theory}
\label{ap:NC}

In this appendix, we provide a formalisation of noncontextual state discrimination, paving the way for the comparison with the quantum model.

\subsection{Ontic space division and noncontextuality}

\begin{figure}
    \centering
    \includegraphics[scale=1.]{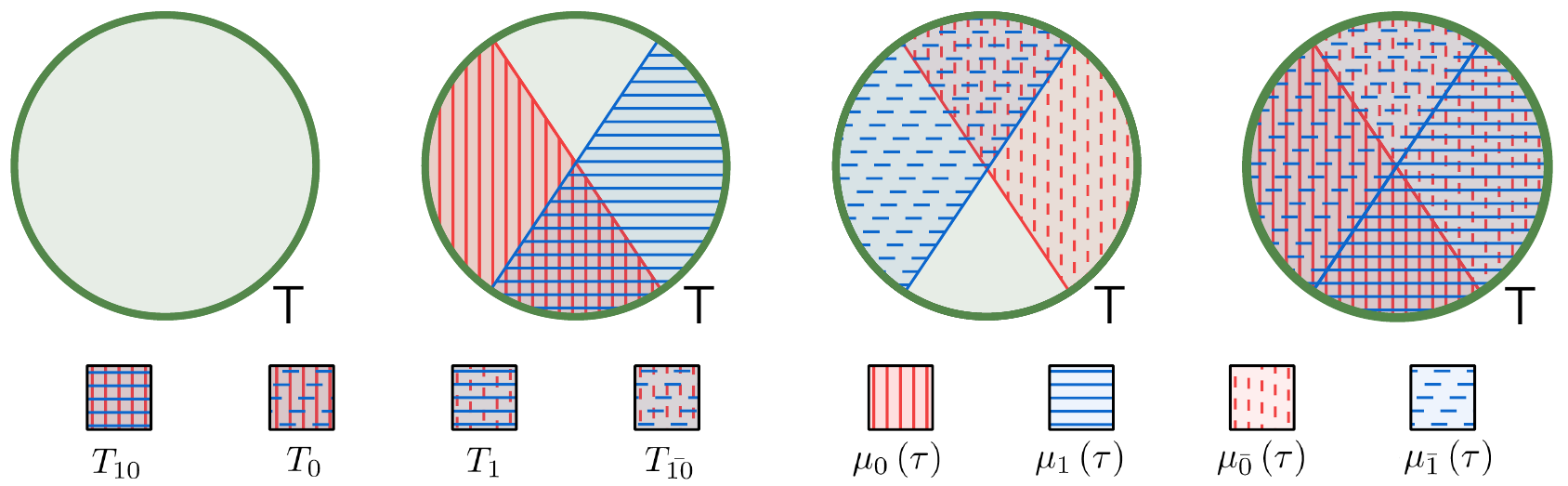}
    \caption{Regions on the ontic space ($T$) according to the overlap of a couple of epistemic states. On the second circle starting form the left, the supports of $\mu_{0}\left(\tau\right)$ and $\mu_{1}\left(\tau\right)$ are drawn. On the third, the supports of $\mu_{\bar{0}}\left(\tau\right)$ and $\mu_{\bar{1}}\left(\tau\right)$ are sketched. On the last circle, the supports of all states are drawn.}
    \label{fig:ontic}
\end{figure}

The observed data in state discrimination problems are the conditional input-output probabilities. In a noncontextual framework, these take the a form analogous to the Born rule in quantum mechanics, see \eqref{eq.ontprob}. In order to simplify the noncontextual optimisation problem, our goal now is to split up the integral over four different regions, as sketched in \figref{fig:ontic}. 

For each epistemic state ($\mu_{x}\left(\tau\right)$) we define its complementary epistemic state($\mu_{\bar{x}}\left(\tau\right)$) which fulfil the orthogonality relation, $\mu_{x}\left(\tau\right)\cdot\mu_{\bar{x}}\left(\tau\right)=0$. Also, the preparation noncontextuality assumption implies that each pair of complementary epistemic states sum to the \textit{maximally mixed state} ($\mu_{\frac{\mathds{1}}{2}}\left(\tau\right)$):
\begin{equation}
\label{eq:maxmix}
    \frac{1}{2}\mu_{0}\left(\tau\right) + \frac{1}{2}\mu_{\bar{0}}\left(\tau\right) = \frac{1}{2}\mu_{1}\left(\tau\right) + \frac{1}{2}\mu_{\bar{1}}\left(\tau\right) = \mu_{\frac{\mathds{1}}{2}}\left(\tau\right) \ .
\end{equation}
The maximally mixed state is introduced within noncontextual models analogously to the quantum maximally mixed state \cite{Schmid2018}.

Let us divide the ontic space in four regions on the ontic space (\figref{fig:ontic}). In each region at least two epistemic states will overlap. For example, $\mu_{0}\left(\tau\right)$ and $\mu_{1}\left(\tau\right)$ will overlap if $\tau\in T_{01}$; or $\mu_{0}\left(\tau\right)$ and $\mu_{\bar{1}}\left(\tau\right)$ overlap if $\tau\in T_{0}$. On the region where two epistemic states overlap they are equal, due to the noncontextuality assumption. Thus, since $\mu_{x}\left(\tau\right)$ and $\mu_{\bar{x}}\left(\tau\right)$ have disjoint supports:
\begin{equation}
\label{eq:noncontext}
    \mu_{0}\left(\tau\right) = \mu_{\bar{0}}\left(\tau\right) = \mu_{1}\left(\tau\right) = \mu_{\bar{1}}\left(\tau\right) = 2\mu_{\frac{\mathds{1}}{2}}\left(\tau\right) \ .
\end{equation}

\subsection{Noncontextual Matrix notation}

When optimising the noncontextual guessing probability, it is not be convenient to work directly with the response functions $\xi_{b}\left(\tau\right)$ and epistemic states $\mu_{x}\left(\tau\right)$. We can reduce the problem to depend on a finite number of real optimisation variables. Let us introduce the following quantities based on integrating the response functions over the regions of the ontic space previously defined:
\begin{equation}
\label{eq:lconst1}
\begin{array}{cclcl}
\alpha_{0b} & = &  \displaystyle\frac{1}{1-\Delta}\int_{T_{0}} d\tau \xi_{b}\left(\tau\right)\mu_{0}\left(\tau\right) & \underset{\mathrm{n.c.}}{=} &  \displaystyle\frac{1}{1-\Delta}\int_{T_{0}} d\tau \xi_{b}\left(\tau\right)\mu_{\bar{1}}\left(\tau\right) \\ \\
\alpha_{1b} & = &  \displaystyle\frac{1}{1-\Delta}\int_{T_{1}} d\tau \xi_{b}\left(\tau\right)\mu_{1}\left(\tau\right) & \underset{\mathrm{n.c.}}{=} &  \displaystyle\frac{1}{1-\Delta}\int_{T_{1}} d\tau \xi_{b}\left(\tau\right)\mu_{\bar{0}}\left(\tau\right) \\ \\
\beta_{b} & = &  \displaystyle\frac{1}{\Delta}\int_{T_{10}} d\tau \xi_{b}\left(\tau\right)\mu_{0}\left(\tau\right) & \underset{\mathrm{n.c.}}{=} &  \displaystyle\frac{1}{\Delta}\int_{T_{10}} d\tau \xi_{b}\left(\tau\right)\mu_{1}\left(\tau\right)\\ \\
\bar{\beta}_{b} & = &  \displaystyle\frac{1}{\Delta}\int_{T_{\bar{10}}} d\tau \xi_{b}\left(\tau\right)\mu_{\bar{0}}\left(\tau\right) & \underset{\mathrm{n.c.}}{=} &  \displaystyle\frac{1}{\Delta}\int_{T_{\bar{10}}} d\tau \xi_{b}\left(\tau\right)\mu_{\bar{1}}\left(\tau\right) \ .
\end{array}
\end{equation}
The second equality in each row of \eqref{eq:lconst1} is fulfilled when preparation noncontextuality is fulfilled, i.e. \eqref{eq:noncontext}. In fact, we can express these terms in a more compact form
\begin{equation}
\label{eq:lconst}
\begin{split}
\alpha_{xb} & =  \displaystyle\frac{2}{1-\Delta}\int_{T_{x}} d\tau \xi_{b}\left(\tau\right)\mu_{\frac{\mathds{1}}{2}}\left(\tau\right)  , \\
\beta_{b} & =  \displaystyle\frac{2}{\Delta}\int_{T_{10}} d\tau \xi_{b}\left(\tau\right)\mu_{\frac{\mathds{1}}{2}}\left(\tau\right) , \\
\bar{\beta}_{b} & =  \displaystyle\frac{2}{\Delta}\int_{T_{\bar{10}}} d\tau \xi_{b}\left(\tau\right)\mu_{\frac{\mathds{1}}{2}}\left(\tau\right)  \ .
\end{split}
\end{equation}
The probabilities in \eqref{eq.ontprob} can be written in terms of these quantities as
\begin{equation}
\label{eq:probsab}
    p\left(b|x\right) = \alpha_{xb}\left(1-\Delta\right)+\beta_{b}\Delta \ .
\end{equation}
It is then sufficient to consider the value of the integration of the response functions times the maximally mixed state, over the regions we introduced, to solve the noncontextual state discrimination problem.

Pushing this notation further, we propose a matrix structure which collects the notion of the divisions of the ontic space in \figref{fig:ontic}. Each epistemic state will be represented by a $2\times 2$ matrix, $\hat{\mu}_{x}$, and each term will represent the definite integral over the different regions on the ontic space, as
\begin{equation}
\label{eq:matep}
\begin{array}{c}
\hat{\mu}_{x}^{T}\equiv\left(\begin{array}{cc}
\displaystyle\int_{T_{0}} d\tau\mu_{x}\left(\tau\right) & \displaystyle\int_{T_{10}} d\tau\mu_{x}\left(\tau\right) \\ \\
\displaystyle\int_{T_{\bar{10}}} d\tau\mu_{x}\left(\tau\right) & \displaystyle\int_{T_{1}} d\tau\mu_{x}\left(\tau\right) \\
\end{array}\right)
=\left(\begin{array}{ll}
\displaystyle\left(\delta_{x,0}+\delta_{x,\bar{1}}\right)\left(1-\Delta\right) & \hspace{0.1cm} \displaystyle\left(\delta_{x,0}+\delta_{x,1}\right)\Delta \\ \\
\displaystyle\left(\delta_{x,\bar{0}}+\delta_{x,\bar{1}}\right)\Delta & \hspace{-0.8cm} \displaystyle\left(\delta_{x,\bar{0}}+\delta_{x,1}\right)\left(1-\Delta\right)
\end{array}\right)
\end{array} \ .
\end{equation}

It is convenient to define the transpose of the matrix form of the epistemic state to ease the notation later on. The orthogonality relation between the complementary and the prepared epistemic states becomes $\hat{\mu}_{x}\circ \hat{\mu}_{\bar{x}}=0$, where $\circ$ is the element-wise matrix product, commonly known as \textit{Hadamard product}, and the right-hand side is the zero matrix. 

We can use the quantities introduced in \eqref{eq:lconst} to write down the matrix form of the response functions, as
\begin{equation}
\label{eq:matresp}
\hat{\xi}_{b}\equiv\left(\begin{array}{cc}
\alpha_{0b} & \beta_{b} \\ \\
\bar{\beta}_{b} & \alpha_{1b} \\
\end{array}\right) \ .
\end{equation} \\
The input-output conditional probabilities can thus be written with the form
\begin{equation}
\label{eq:matrprob}
p\left(b|M,x\right) = \sum_{ij}^{N}\left[\hat{\xi}_{b}\circ\hat{\mu}_{x}^{T}\right]_{ij} = \mathrm{Tr}\left[\hat{\xi}_{b}\hat{\mu}_{x}\right] \ .
\end{equation}

The first equality can be derived straight from \eqref{eq:probsab}, by summing up all the terms from the Hadamard product between the response function and epistemic state. The second equality holds for any pair of $N\times N$ matrices, relating the Hadamard product with the usual matrix product. The result from \eqref{eq:matrprob} allows us to write the input-output probabilities on a form similar to the Born rule in quantum mechanics.

\begin{figure*}
    \centering
    \includegraphics[scale=0.8]{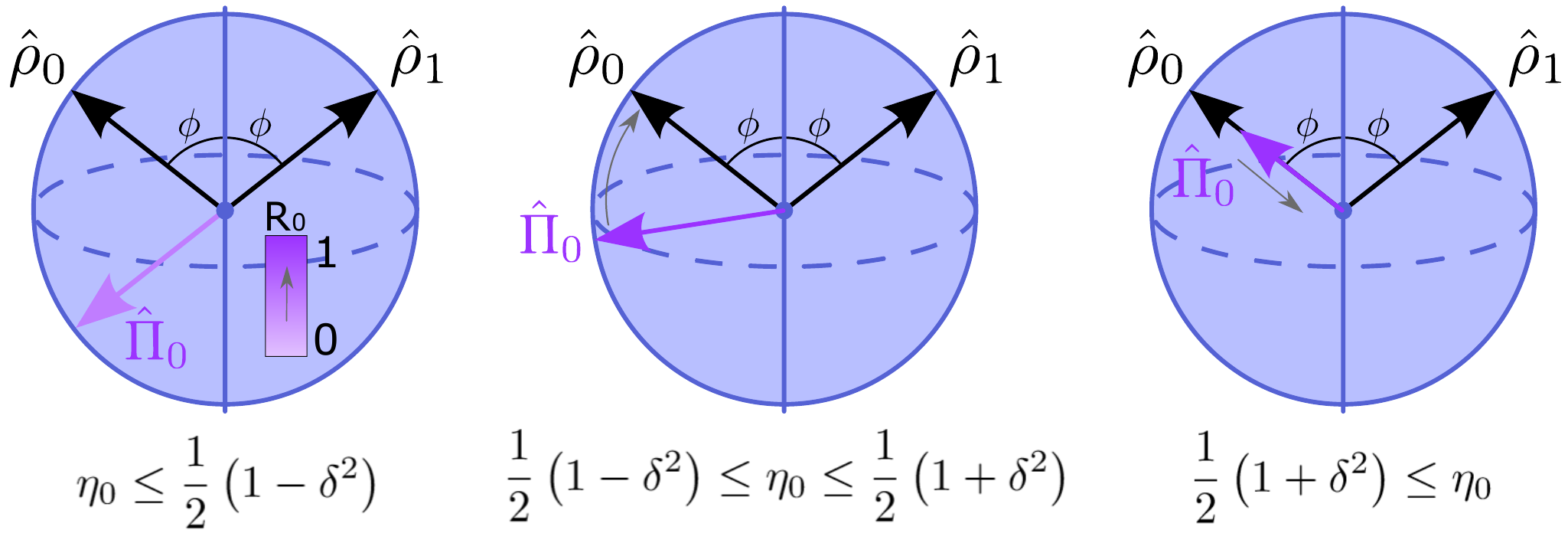} 
    \caption{Representation of the POVM element $\hat{\Pi}_{0}$ and the quantum states $\hat{\rho}_{0}$ and $\hat{\rho}_{1}$ on the Bloch sphere for different values of the output rate $\eta_{0}$.}
    \label{fig:Bloch_Sphere}
\end{figure*}

\section{Rates at which states are unambiguously identified}
\label{ap:USDinMCD}

In this appendix we derive analytical expressions for the output rates at which optimal MCD unambiguously identifies one of the inputs (solid vertical lines in \figref{fig.Hmincompare}). We will look separately at the quantum and noncontextual scenarios.

\subsection{Quantum case}

We look at the discrimination of two pure states $\hat{\rho}_{0}=\ket{\psi_{0}}\bra{\psi_{0}}$ and $\hat{\rho}_{1}=\ket{\psi_{1}}\bra{\psi_{1}}$ randomly prepared with equal probability. The POVM element corresponding to outcome $b$ can be represented on the Bloch sphere as
\begin{equation}
\label{eq:POVM0}
    \hat{\Pi}_{b} = \frac{R_b}{2}\left[\mathds{1}+r_b\sin\theta_b\hat{\sigma}_{x}+r_b\cos\theta_b\hat{\sigma}_z\right]\ ,
\end{equation}
where $\hat{\sigma}_{x}$ and $\hat{\sigma}_{z}$ are the Pauli matrices,  $\left(r_b\sin\theta_b,0,r_b\cos\theta_b\right)$ is the Bloch vector (on the X-Z plane), with $|r_b|\leq 1$, $R_b\geq 0$, and $\left\{\hat{\sigma}_x,\hat{\sigma}_z\right\}$ are the Pauli matrices \cite{Tavakoli2020sciadv}. Positivity and normalisation imply that
\begin{equation}
\label{eq:norm}
    \sum_{b}R_b=2, \ \sum_{b}R_b r_b \sin\theta_b = \sum_{b}R_b r_b \cos\theta_b = 0 \ .
\end{equation} 
In \figref{fig:Bloch_Sphere} we show the POVM element $\hat{\Pi}_{0}$ in the Bloch sphere, together with the quantum states
\begin{equation}
\begin{split}
    & \hat{\rho}_{0}=\frac{1}{2}\left[\mathds{1}-\sin\phi \, \hat{\sigma}_{x}+\cos\phi \, \hat{\sigma}_{z}\right] , \\
    & \hat{\rho}_{1}=\frac{1}{2}\left[\mathds{1}+\sin\phi \, \hat{\sigma}_{x}+\cos\phi \, \hat{\sigma}_{z}\right]  \ .
\end{split}    
\end{equation}
The overlap is given by $\cos\phi=\delta$. The confidence $C_{0}$ expressed in terms of $\hat{\Pi}_{0}$ is
\begin{align}
\label{eq:C0POVM}
\hspace{-0.2cm}
    C_{0} = \frac{\mathrm{Tr}\left[\hat{\Pi}_{0}\hat{\rho}_{0}\right]}{\mathrm{Tr}\left[\hat{\Pi}_{0}\hat{\rho}_{0}\right]+\mathrm{Tr}\left[\hat{\Pi}_{0}\hat{\rho}_{1}\right]} .
\end{align}
The expression in \eqref{eq:C0POVM} is the figure of merit in MCD. Without loos of generality, we can focus on the POVM element $\hat{\Pi}_{0}$, and consider $\hat{\Pi}_{1}=\hat{\Pi}_{\o}$.

The maximum value of the confidence ($C_{0}=1$) can be obtained if the measurement device is able to unambiguously discriminate the state $\hat{\rho}_{0}$, i.e.  $\mathrm{Tr}\left[\hat{\Pi}_{0}\hat{\rho}_{1}\right]=0$. This implies that $\theta_0=\phi+\pi$ and $r_0=1$. The POVM element $\hat{\Pi}_{0}$ has rank $1$, and we are left with $0\leq R_{0}\leq 1$. The only possible output rates are
\begin{equation}
    0\leq \eta_{0} \leq \frac{1}{2}\left(1-\delta^{2}\right)\ .
\end{equation}
For higher rates, we need to allow $\hat{\Pi}_{0}$ to rotate.  To keep $C_{0}$ as large as possible, we need to make sure that the numerator is also at its maximum. Thus, our goal now is to find the maximum value of $p(0|0)$. That is achieved by rotating $\hat{\Pi}_{0}$ towards $\hat{\rho}_{0}$. The rotation angle $\theta_0$ can be parametrized in terms of the output rate as, $\cos\theta_0 = (2\eta_{0}-1)/\delta$. This will run within the interval $\phi+\pi \leq \theta \leq 2\pi-\phi$. For the output rate, this means
\begin{equation}
    \frac{1}{2}\left(1-\delta^{2}\right) \leq \eta_{0} \leq \frac{1}{2}\left(1+\delta^{2}\right) \ .
\end{equation}
Beyond that point, the output rate saturates when the POVM element $\hat{\Pi}_{0}$ is no longer projective. Thus, $r_{0}$ will be reduced to zero, while keeping $R_0=1$. For the output rate this means
\begin{equation}
    \frac{1}{2}\left(1+\delta^{2}\right)\leq\eta_{0}\leq 1 \ .
\end{equation}
Here state $\hat{\rho}_{1}$ is unambiguously discriminated, i.e.  $\mathrm{Tr}\left[\hat{\Pi}_{1}\hat{\rho}_{0}\right]=0$, as $\theta_{1}=\theta_{ø}=\pi-\phi$ becuase of \eqref{eq:norm}.

\subsection{Noncontextual case}

In the noncontextual framework, we use \eqref{eq:lconst}, and the probabilities \eqref{eq:probsab}. Then, the confidence can be written as
\begin{equation}
C_{0}  = \frac{\left(\alpha_{00}\left(1-\Delta\right) + \beta_{0}\Delta\right)}{\left(\alpha_{00}+\alpha_{10}\right)\left(1-\Delta\right) + 2\beta_{0}\Delta}
\end{equation}
The maximal value on the confidence ($C_{0}=1$) is achieved when $\alpha_{10}=\beta_{0}=0$. Since $0\leq \alpha_{00} \leq 1$, this occurs for rates
\begin{equation}
0\leq \eta_{0}\leq \frac{1}{2}\left(1-\Delta\right) \ .
\end{equation}
For larger rates, we need $\beta_{0}$ to grow. We can keep $\alpha_{10}=0$ since it only appears in the denominator. Again, since $0\leq \beta_{0} \leq 1$, the rates at which this is possible are
\begin{equation}
\frac{1}{2}\left(1-\Delta\right)\leq \eta_{0}\leq \frac{1}{2}\left(1+\Delta\right)
\end{equation}
Finally, for even larger $\eta_{0}$ we need $\alpha_{10}$ to grow. As $0\leq \alpha_{10} \leq 1$, we are left with
\begin{equation}
\frac{1}{2}\left(1+\Delta\right)\leq \eta_{0}\leq 1 \ .
\end{equation}



\section{SDP for noncontextual randomness certification}
\label{ap:RNG_NC}
In this appendix, we show that the average guessing probability for a noncontextual eavesdropper (in \eqref{eq.NCopt}) can be cast as an SDP, similarly to the quantum case. The derivation closely follows \appref{ap:QRNG}.

As in \appref{ap:QRNG}, we the number of relevant strategies is again $9$, labeled by $\lambda_{x}\in\left\{0,1,ø\right\}$ for $x\in\left\{0,1\right\}$.  These distribution over strategies and the response functions fulfill
\begin{align}
\sum_{\lambda_{0},\lambda_{1}}q_{\lambda_{0}\lambda_{1}} &=1 , \\ 
q_{\lambda_{0}\lambda_{1}} & \geq 0 \,\, \forall \lambda_{0},\lambda_{1} , \\
\sum_{b}\xi_{b}^{\lambda_{0}\lambda_{1}}(\tau) & = 1 \,\, \forall \lambda_{0},\lambda_{1},\tau , \\
\xi_{b}^{\lambda_{0}\lambda_{1}}(\tau) & \geq 0 \,\, \forall \lambda_{0},\lambda_{1},b,\tau  ,
\end{align}
and the observed output rate $\eta_{0}$ and confidence $C_{0}^{NC}$ should be reproduced,
\begin{align}
\sum_{\lambda_{0},\lambda_{1}}\sum_{x}p_{x}q_{\lambda_{0}\lambda_{1}}\int d\tau\xi^{\lambda_{0}\lambda_{1}}_{0}(\tau)\mu_{x}(\tau) & = \eta_{0} \label{eq:e1c}\\
\sum_{\lambda_{0},\lambda_{1}}\frac{p_{0}}{\eta_{0}}q_{\lambda_{0}\lambda_{1}}\int d\tau\xi^{\lambda_{0}\lambda_{1}}_{0}(\tau)\mu_{0}(\tau) & = C_{0}^{Q} \label{eq:e1d} \ .
\end{align}
Finally, $p_{g}^{NC}$ can be linearised by defining $M_{b}^{\lambda_{0}\lambda_{1}}(\tau)=q_{\lambda_{0}\lambda_{1}}\xi_{b}^{\lambda_{0}\lambda_{1}}(\tau)$. The primal version of the SDP can then be written as follows:
\begin{equation}
    \label{eq:primNC}
    \boxed{
    \begin{array}{l}
    \begin{array}{ll}
      \underset{M_{b}^{\lambda_{0}\lambda_{1}}(\tau)}{\mathrm{maximise}}  & \displaystyle p_{g}^{NC} = \sum_{x=0}^{1}\sum_{\lambda_{0},\lambda_{1}}p_{x}\int d\tau M^{\lambda_{0}\lambda_{1}}_{\lambda_{x}}(\tau)\mu_{x}(\tau) \\ \\
      \mathrm{subject \ to:}  & 
     \end{array} \\ \\
     \begin{array}{l}
       \displaystyle M_{b}^{\lambda_{0}\lambda_{1}}(\tau)\geq 0, \ \forall \lambda_{0},\lambda_{1},b,\tau \\ \\
       \displaystyle\sum_{b} M_{b}^{\lambda_{0}\lambda_{1}}(\tau) = \frac{1}{|T|}\int d\tau\sum_{b}M_{b}^{\lambda_{0}\lambda_{1}}(\tau), \ \forall \lambda_{0},\lambda_{1} \\ \\
      \displaystyle  \sum_{b}\sum_{\lambda_{0},\lambda_{1}}\sum_{x}p_{x}\int d\tau M^{\lambda_{0}\lambda_{1}}_{b}(\tau)\mu_{x}(\tau) = 1 \\ \\
      \displaystyle\sum_{\lambda_{0},\lambda_{1}}\sum_{x}p_{x}\int d\tau M^{\lambda_{0}\lambda_{1}}_{0}(\tau)\mu_{x}(\tau) = \eta_{0}  \\ \\
      \displaystyle\sum_{\lambda_{0},\lambda_{1}}\frac{p_{0}}{\eta_{0}}\int d\tau M^{\lambda_{0}\lambda_{1}}_{0}(\tau)\mu_{0}(\tau) = C_{0}^{NC} \ .
    \end{array}
    \end{array}}
\end{equation}
The last two constraints can be reduced to:
\begin{align}
\label{eq:comb}
    \sum_{\lambda_{0},\lambda_{1}}p_{x}\int d\tau M^{\lambda_{0}\lambda_{1}}_{0}(\tau)\mu_{x}(\tau)=\eta_{0}C_{0}^{NC}\delta_{x,0} + \eta_{0}\left(1-C_{0}^{NC}\right)\delta_{x,1} \nonumber
\end{align}
The explicit use of functions over the ontic space as optimisation variables and the presence of integrals makes th SDP impractical to solve. To avoid these issues, we introduce the quantities from \eqref{eq:lconst} and define:
\begin{equation}
\label{eq:llconst}
\begin{split}
A_{xb}^{\lambda_{0}\lambda_{1}} & = q_{\lambda_{0}\lambda_{1}}\alpha_{xb} =  \displaystyle\frac{2}{1-\Delta}\int_{T_{x}} d\tau M_{b}^{\lambda_{0}\lambda_{1}}\left(\tau\right)\mu_{\frac{\mathds{1}}{2}}\left(\tau\right) , \\
B_{b}^{\lambda_{0}\lambda_{1}} & = q_{\lambda_{0}\lambda_{1}}\beta_{b} =  \displaystyle\frac{2}{\Delta}\int_{T_{10}} d\tau M_{b}^{\lambda_{0}\lambda_{1}}\left(\tau\right)\mu_{\frac{\mathds{1}}{2}}\left(\tau\right) , \\
\bar{B}_{b}^{\lambda_{0}\lambda_{1}} & = q_{\lambda_{0}\lambda_{1}}\bar{\beta}_{b} = \displaystyle\frac{2}{\Delta}\int_{T_{\bar{10}}} d\tau M_{b}^{\lambda_{0}\lambda_{1}}\left(\tau\right)\mu_{\frac{\mathds{1}}{2}}\left(\tau\right) \ .
\end{split}
\end{equation}
We can now re-write the primal problem in \eqref{eq:primNC} with the quantities in \eqref{eq:llconst}. Since this process in not trivial, we go through it step by step.
\begin{itemize}
    \item Guessing probability:
    \begin{align}
       & p_{g}^{NC} = \sum_{x}\sum_{\lambda_{0},\lambda_{1}}p_{x} \\
       & \left[\int_{T_{0}} d\tau M_{\lambda_{x}}^{\lambda_{0}\lambda_{1}}\left(\tau\right)\mu_{x}\left(\tau\right) + \int_{T_{10}} d\tau M_{\lambda_{x}}^{\lambda_{0}\lambda_{1}}\left(\tau\right)\mu_{x}\left(\tau\right)\right. + \left.\int_{T_{1}} d\tau M_{\lambda_{x}}^{\lambda_{0}\lambda_{1}}\left(\tau\right)\mu_{x}\left(\tau\right) + \int_{T_{\bar{10}}} d\tau M_{\lambda_{x}}^{\lambda_{0}\lambda_{1}}\left(\tau\right)\mu_{x}\left(\tau\right)\right] \nonumber \\
       &  = \sum_{x}\sum_{\lambda_{0},\lambda_{1}}p_{x}\left[\left(1-\Delta\right)A_{x\lambda_{x}}^{\lambda_{0}\lambda_{1}} + \Delta B_{\lambda_{x}}^{\lambda_{0}\lambda_{1}} \right] \nonumber
    \end{align}
\end{itemize}
\begin{itemize}
    \item Non-negativity constraint:
    \begin{equation}
    M_{b}^{\lambda_{0}\lambda_{1}}\left(\tau\right)\geq  0 \Leftrightarrow \left\{\begin{array}{l}
    A_{xb}^{\lambda_{0}\lambda_{1}}\geq 0 \\ \\
    B_{b}^{\lambda_{0}\lambda_{1}}\geq 0 \\ \\
    \bar{B}_{b}^{\lambda_{0}\lambda_{1}}\geq 0
    \end{array}\right. \forall \lambda_{0},\lambda_{1},x,b,
    \end{equation}
    
    \item Ontic state independence from $q_{\lambda_{0}\lambda_{1}}$:
    \begin{equation}
    \label{eq:constA}
    \begin{array}{l}
        \displaystyle \sum_{b}A_{xb}^{\lambda_{0}\lambda_{1}} = \frac{1}{1-\Delta}\int_{T_{x}}d\tau \sum_{b}M_{b}^{\lambda_{0}\lambda_{1}}\left(\tau\right) \mu_{x}\left(\tau\right)
        =\displaystyle \sum_{b}M_{b}^{\lambda_{0}\lambda_{1}}\left(\tau'\right) \frac{1}{1-\Delta}\int_{T_{x}}d\tau\mu_{x}\left(\tau\right) =\displaystyle\sum_{b}M_{b}^{\lambda_{0}\lambda_{1}}\left(\tau'\right) \ .
    \end{array}
    \end{equation}
    Also:
    \begin{equation}
    \label{eq:constB}
        \sum_{b}B_{b}^{\lambda_{0}\lambda_{1}} = \sum_{b}\bar{B}_{b}^{\lambda_{0}\lambda_{1}} =\sum_{b}M_{b}^{\lambda_{0}\lambda_{1}}\left(\tau'\right) \ .
    \end{equation}
    On the other hand:
    \begin{align}
    \label{eq:constC}
        &\sum_{b}\left[\left(1-\Delta\right)\left(A_{0b}^{\lambda_{0}\lambda_{1}}+A_{1b}^{\lambda_{0}\lambda_{1}}\right) \Delta\left(B_{b}^{\lambda_{0}\lambda_{1}}+\bar{B}_{b}^{\lambda_{0}\lambda_{1}}\right)\right] \\
        = &\sum_{b}\left[\int_{T_{0}} d\tau M_{b}^{\lambda_{0}\lambda_{1}}\left(\tau\right)\mu_{\frac{\mathds{1}}{2}}\left(\tau\right)
        + \int_{T_{1}} d\tau M_{b}^{\lambda_{0}\lambda_{1}}\left(\tau\right)\mu_{\frac{\mathds{1}}{2}}\left(\tau\right) 
        + \int_{T_{10}} d\tau M_{b}^{\lambda_{0}\lambda_{1}}\left(\tau\right)\mu_{\frac{\mathds{1}}{2}}\left(\tau\right)
        + \int_{T_{\bar{10}}} d\tau M_{b}^{\lambda_{0}\lambda_{1}}\left(\tau\right)\mu_{\frac{\mathds{1}}{2}}\left(\tau\right)\right] \nonumber \\
        = &\sum_{b}M_{b}^{\lambda_{0}\lambda_{1}}\left(\tau'\right)\int_{T}d\tau 2\mu_{\frac{\mathds{1}}{2}}\left(\tau\right) = 2\sum_{b}M_{b}^{\lambda_{0}\lambda_{1}}\left(\tau'\right) \nonumber
    \end{align}
    Thus, combining \eqref{eq:constA}, \eqref{eq:constB} and \eqref{eq:constC} one ends up with:
    \begin{align}
    &\sum_{b}A_{xb}^{\lambda_{0}\lambda_{1}} = \sum_{b}B_{b}^{\lambda_{0}\lambda_{1}} = \sum_{b}\bar{B}_{b}^{\lambda_{0}\lambda_{1}} = \sum_{b}\left[\frac{1-\Delta}{2}\left(A_{0b}^{\lambda_{0}\lambda_{1}}+A_{1b}^{\lambda_{0}\lambda_{1}}\right)+ \frac{\Delta}{2}\left(B_{b}^{\lambda_{0}\lambda_{1}}+\bar{B}_{b}^{\lambda_{0}\lambda_{1}}\right)\right] \nonumber
    \end{align}
    \item Reproduce the output rates:
    \begin{align}
        \eta_{b} = & \sum_{\lambda_{0}\lambda_{1}}\sum_{x}p_{x} \\
        &\left[\int_{T_{0}}d\tau M_{b}^{\lambda_{0}\lambda_{1}}\left(\tau\right)\mu_{x}\left(\tau\right) + \int_{T_{10}}d\tau M_{b}^{\lambda_{0}\lambda_{1}}\left(\tau\right)\mu_{x}\left(\tau\right)+\int_{T_{1}}d\tau M_{b}^{\lambda_{0}\lambda_{1}}\left(\tau\right)\mu_{x}\left(\tau\right) + \int_{T_{\bar{10}}}d\tau M_{b}^{\lambda_{0}\lambda_{1}}\left(\tau\right)\mu_{x}\left(\tau\right)\right] \nonumber \\
        = &\sum_{\lambda_{0}\lambda_{1}}\sum_{x}p_{x}\left[\left(1-\Delta\right)A_{xb}^{\lambda_{0}\lambda_{1}} + \Delta B_{b}^{\lambda_{0}\lambda_{1}}\right] \nonumber
    \end{align}
    \item Normalisation of the output rates:
    \begin{align}
        &\sum_{b}\sum_{\lambda_{0}\lambda_{1}}\sum_{x}p_{x} \\
        &\left[\int_{T_{0}}d\tau M_{b}^{\lambda_{0}\lambda_{1}}\left(\tau\right)\mu_{x}\left(\tau\right) + \int_{T_{10}}d\tau M_{b}^{\lambda_{0}\lambda_{1}}\left(\tau\right)\mu_{x}\left(\tau\right)+\int_{T_{1}}d\tau M_{b}^{\lambda_{0}\lambda_{1}}\left(\tau\right)\mu_{x}\left(\tau\right) + \int_{T_{\bar{10}}}d\tau M_{b}^{\lambda_{0}\lambda_{1}}\left(\tau\right)\mu_{x}\left(\tau\right)\right] \nonumber \\
        = &\sum_{b}\sum_{\lambda_{0}\lambda_{1}}\sum_{x}p_{x}\left[\left(1-\Delta\right)A_{xb}^{\lambda_{0}\lambda_{1}} + \Delta B_{b}^{\lambda_{0}\lambda_{1}}\right]=1 \nonumber
    \end{align}
    
    \item Fix the confidence $C_{0}$ of the measurement device:
    \begin{align}
        &\sum_{\lambda_{0}\lambda_{1}}\frac{p_{0}}{\eta_{0}}\int_{T}d\tau M_{0}^{\lambda_{0}\lambda_{1}}\left(\tau\right)\mu_{0}\left(\tau\right) = \sum_{\lambda_{0}\lambda_{1}}\frac{p_{0}}{\eta_{0}}\left[\left(1-\Delta\right)A_{0b}^{\lambda_{0}\lambda_{1}} + \Delta B_{b}^{\lambda_{0}\lambda_{1}}\right]=C_{0}^{NC} \nonumber
    \end{align}

\end{itemize}
At the end of the day, we can write the primal problem as follows:

\begin{equation}
    \label{eq:primNC}
    \boxed{
    \begin{array}{l}
    \begin{array}{ll}
      \underset{\left\{A_{xb}^{\lambda_{0}\lambda_{1}},B_{b}^{\lambda_{0}\lambda_{1}},\bar{B}_{b}^{\lambda_{0}\lambda_{1}}\right\}}{\mathrm{maximise}}  & \displaystyle p_{g}^{NC} = \sum_{x}\sum_{\lambda_{0},\lambda_{1}}p_{x}\left[\left(1-\Delta\right)A_{x\lambda_{x}}^{\lambda_{0}\lambda_{1}} + \Delta B_{\lambda_{x}}^{\lambda_{0}\lambda_{1}} \right]  \\ \\
      \mathrm{subject \ to:}  & \displaystyle A_{xb}^{\lambda_{0}\lambda_{1}}\geq 0,B_{b}^{\lambda_{0}\lambda_{1}} \geq 0,\bar{B}_{b}^{\lambda_{0}\lambda_{1}} \geq 0 \ \forall \lambda_{0},\lambda_{1},b
     \end{array} \\ \\
     \begin{array}{l}
       \begin{array}{l}
    \displaystyle\sum_{b}A_{0b}^{\lambda_{0}\lambda_{1}} = \sum_{b}A_{1b}^{\lambda_{0}\lambda_{1}} = \sum_{b}B_{b}^{\lambda_{0}\lambda_{1}} = \sum_{b}\bar{B}_{b}^{\lambda_{0}\lambda_{1}} = \\
    \displaystyle = \sum_{b}\left[\frac{1-\Delta}{2}\left(A_{0b}^{\lambda_{0}\lambda_{1}}+A_{1b}^{\lambda_{0}\lambda_{1}}\right)+ \frac{\Delta}{2}\left(B_{b}^{\lambda_{0}\lambda_{1}}+\bar{B}_{b}^{\lambda_{0}\lambda_{1}}\right)\right]
    \end{array} \\ \\
    \displaystyle\sum_{b}\sum_{\lambda_{0}\lambda_{1}}\sum_{x}p_{x}\left[\left(1-\Delta\right)A_{xb}^{\lambda_{0}\lambda_{1}} + \Delta B_{b}^{\lambda_{0}\lambda_{1}}\right] =1 \\ \\
      \displaystyle \sum_{\lambda_{0}\lambda_{1}}\sum_{x}p_{x}\left[\left(1-\Delta\right)A_{x0}^{\lambda_{0}\lambda_{1}} + \Delta B_{0}^{\lambda_{0}\lambda_{1}}\right]=\eta_{0} \\ \\
      \displaystyle\sum_{\lambda_{0}\lambda_{1}}\frac{p_{0}}{\eta_{0}}\left[\left(1-\Delta\right)A_{0b}^{\lambda_{0}\lambda_{1}} + \Delta B_{b}^{\lambda_{0}\lambda_{1}}\right]=C_{0}^{NC} \ .
    \end{array}
    \end{array}}
\end{equation}

Using the matrix form of the response functions introduced in \eqref{eq:matresp}, we define:
\begin{equation}
\label{eq:matopt}
\hat{M}_{b}^{\lambda_{0}\lambda_{1}}\equiv q_{\lambda_{0}\lambda_{1}}\hat{\xi}_{b}^{\lambda_{0}\lambda_{1}}=\left(\begin{array}{cc}
A_{1b}^{\lambda_{0}\lambda_{1}} & B_{b}^{\lambda_{0}\lambda_{1}} \\ \\
\bar{B}_{b}^{\lambda_{0}\lambda_{1}} & A_{2b}^{\lambda_{0}\lambda_{1}} \\
\end{array}\right) \ .
\end{equation} 
Implementing this matrix notation, together with the matrix form of the epistemic states in \eqref{eq:matep}, we re-write the primal problem in \eqref{eq:primNC} as follows:
\begin{equation}
    \label{eq:primNCm}
    \hspace{-0.3cm}\boxed{
    \begin{array}{l}
    \begin{array}{ll}
      \underset{\hat{M}_{b}^{\lambda_{0}\lambda_{1}}}{\mathrm{maximise}}  & \displaystyle p_{g}^{NC} = \sum_{x}\sum_{\lambda_{0},\lambda_{1}}p_{x}\mathrm{Tr}\left[\hat{M}^{\lambda_{0}\lambda_{1}}_{\lambda_{x}}\hat{\mu}_{x}\right]  \\ \\
      \mathrm{subject \ to:}  & \displaystyle\hat{M}_{b}^{\lambda_{0}\lambda_{1}}\underset{\text{e.w.}}{\geq} 0 \ \forall \lambda_{0},\lambda_{1},b
     \end{array} \\ \\
     \begin{array}{l}
       \displaystyle\sum_{b}\hat{M}_{b}^{\lambda_{0}\lambda_{1}} = \mathrm{Tr}\left[\sum_{b}\hat{M}_{b}^{\lambda_{0}\lambda_{1}}\hat{\mu}_{\frac{\mathds{1}}{2}}\right]\hat{J}_{2} \\ \\
      \displaystyle \sum_{b}\sum_{\lambda_{0},\lambda_{1}}\sum_{x}p_{x}\mathrm{Tr}\left[\hat{M}^{\lambda_{0}\lambda_{1}}_{b}\hat{\mu}_{x}\right] = 1 \\ \\
      \displaystyle \sum_{\lambda_{0},\lambda_{1}}\sum_{x}p_{x}\mathrm{Tr}\left[\hat{M}^{\lambda_{0}\lambda_{1}}_{0}\hat{\mu}_{x}\right]=\eta_{0} \\ \\
      \displaystyle\sum_{\lambda_{0},\lambda_{1}}\frac{p_{0}}{\eta_{0}}\mathrm{Tr}\left[\hat{M}^{\lambda_{0}\lambda_{1}}_{0}\hat{\mu}_{0}\right] = C_{0}^{Q} \ .
    \end{array}
    \end{array}}
\end{equation}
Here, $\hat{J}_{2}$ denotes a $2\times2$ matrix with all entries equal to 1. Also, $\underset{\text{e.w.}}{\geq}$ denotes element-wise inequalities between matrices, and the maximally mixed state in noncontextual theory has been introduced in the matrix notation. It is given by
\begin{equation}
\label{eq:matmaxmix}
\hat{\mu}_{\frac{\mathds{1}}{2}}\equiv\frac{1}{2}\left(\begin{array}{cc}
1-\Delta & \Delta \\ \\
\Delta & 1-\Delta \\
\end{array}\right) \ .
\end{equation}
Finally, note that both last constraints in \eqref{eq:primNCm} can be re-written as:
\begin{equation}
\label{eq:combNC}
    \sum_{\lambda_{0},\lambda_{1}}p_{x}\mathrm{Tr}\left[\hat{M}^{\lambda_{0}\lambda_{1}}_{0}\hat{\mu}_{x}^{T}\right] = C_{0}^{NC}\delta_{x,0} + \left(1-C_{0}^{NC}\right)\delta_{x,1}
\end{equation}
Also, due to normalization:
\begin{equation}
    \sum_{b}\sum_{\lambda_{0},\lambda_{1}}\sum_{x}\mathrm{Tr}\left[\hat{M}^{\lambda_{0}\lambda_{1}}_{b}\hat{\mu}_{x}\right] = 2
\end{equation}
We proceed obtaining the dual problem in the noncontextual framework. From each constraint in \eqref{eq:primNCm}, we introduce the dual variables: $\hat{G}_{b}^{\lambda_{0}\lambda{1}}$, $\hat{H}^{\lambda_{0}\lambda{1}}$, $\nu_{x}$ and $\chi$. The corresponding Lagrangian will then be:
\begin{align}
\mathcal{L} = &\sum_{x}\sum_{\lambda_{0},\lambda_{1}}p_{x}\mathrm{Tr}\left[\hat{\mu}_{x}\hat{M}_{\lambda_{x}}^{\lambda_{0}\lambda_{1}}\right]
+\sum_{b}\sum_{\lambda_{1},\lambda_{2}}\mathrm{Tr}\left[\hat{G}_{b}^{\lambda_{0}\lambda_{1}}\hat{M}_{b}^{\lambda_{0}\lambda_{1}}\right] 
+\sum_{\lambda_{0},\lambda_{1}}\mathrm{Tr}\left[\hat{H}^{\lambda_{0}\lambda_{1}}\sum_{b}\left(\hat{M}_{b}^{\lambda_{0}\lambda_{1}}-\mathrm{Tr}\left[\hat{M}_{b}^{\lambda_{0}\lambda_{1}}\hat{\mu}_{\frac{\mathds{1}}{2}}\right]\hat{J}_{2}\right)\right] \nonumber \\
&+\sum_{x}\nu_{x}\left(\sum_{\lambda_{0},\lambda_{1}}p_{x}\mathrm{Tr}\left[\hat{\mu}_{x}\hat{M}_{0}^{\lambda_{0}\lambda_{1}}\right]-\eta_{0}\left(\delta_{x,0}C_{0}^{NC}+\delta_{x,1}\left(1-C_{0}^{NC}\right)\right)\right)
+\chi\left(\sum_{b}\sum_{\lambda_{0},\lambda_{1}}\sum_{x}p_{x}\mathrm{Tr}\left[\hat{\mu}_{x}\hat{M}_{b}^{\lambda_{1}\lambda_{2}}\right]-1\right) \ \label{eq:lagNC} .
\end{align}
We write the supremum of the Lagrangian as:
\begin{equation}
    \label{eq:supNC}
    \mathcal{S}\equiv\underset{\hat{M}_{b}^{\lambda_{0}\lambda_{1}}}{\mathrm{supp}}  \mathcal{L} \ .
\end{equation}
Given any solution $\hat{M}_{b}^{\lambda_{0}\lambda_{1}}$ of the primal, the last three terms in \eqref{eq:lagNC} vanish. Thus, as $\hat{M}_{b}^{\lambda_{0}\lambda_{1}}$ are constrained to be positive semi-definite, the first line in \eqref{eq:lagQ} yields an upper bound on the guessing probability $p_{g}^{NC}$ (only if all $\hat{G}_{b}^{\lambda_{0}\lambda_{1}}$ are positive semi-definite). The dual can then be formulated by minimising the supremum in \eqref{eq:supNC}. We re-write it as follows:
\begin{align}
\label{eq:sNCr}
    \mathcal{S}=\underset{\hat{M}_{b}^{\lambda_{0}\lambda_{1}}}{\mathrm{supp}} \sum_{\lambda_{0},\lambda_{1}}\mathrm{Tr}\left[\hat{M}_{b}^{\lambda_{0}\lambda_{1}}\hat{K}^{\lambda_{0}\lambda_{1}}_{b}\right]-\sum_{x}\nu_{x}\eta_{0}\left(\delta_{x,0}C_{0}^{NC}+\delta_{x,1}\left(1-C_{0}^{NC}\right)\right)-\chi,
\end{align}
where,
\begin{align}
\hat{K}_{b}^{\lambda_{0}\lambda_{1}}=\sum_{x}p_{x}\hat{\mu}_{x}\left(\delta_{b,\lambda_{x}}+\nu_{x}\delta_{b,0}+\chi\right)+\hat{G}_{b}^{\lambda_{0}\lambda_{1}}+\hat{H}^{\lambda_{0}\lambda_{1}}-\mathrm{Tr}\left[\hat{H}^{\lambda_{0}\lambda_{1}}\hat{J}_{2}\right]\hat{\mu}_{\frac{\mathds{1}}{2}} \ .
\end{align}
The supremum in \eqref{eq:sNCr} will diverge, unless  $\hat{K}_{b}^{\lambda_{0}\lambda_{1}}=0$. We will drop $\hat{G}_{b}^{\lambda_{0}\lambda_{1}}$, imposing that the remaining expression is negative. This way, the guessing probability can be upper bounded by:
\begin{equation}
    p_{g}\leq p_{g}^{NC}= -\sum_{x=0}^{1}\nu_{x}\left(\delta_{x,0}C_{0}^{NC}+\delta_{x,1}\left(1-C_{0}^{NC}\right) \right)-\chi
\end{equation}
for a given value of confidence $C_{0}$ in discriminating $\hat{\rho}_{1}$ and any numbers $\nu_{x}$ and $\chi$ which fulfil that there exists four $2\times 2$ matrices $\hat{H}^{\lambda_{0}\lambda_{1}}$, with indices $\lambda_{0},\lambda_{1}=0,1,ø$, such that:
\begin{align}
\sum_{x=0}^{1}p_{x}\hat{\mu}_{x}\left(\delta_{b,\lambda_{x}}+\nu_{x}\delta_{b,0}+\chi
\right) + \hat{H}^{\lambda_{0}\lambda_{1}}-\mathrm{Tr}\left[\hat{H}^{\lambda_{0}\lambda_{1}}\hat{J}_{2}\right]\hat{\mu}_{\frac{\mathds{1}}{2}} \leq 0 \ .
\end{align}

As a final remark, note that one can straightforwardly switch between quantum and noncontextual models by switching: the bound on the confidence ($C_{0}^{Q}\leftrightarrow C_{0}^{NC}$); the physical state representations ($\hat{\rho}_{x}\leftrightarrow\hat{\mu}_{x}$); the measurement outcome representation ($\hat{\Pi}_{b}^{\lambda}\leftrightarrow\hat{\xi}_{b}^{\lambda}$); the identity element ($\mathds{1}\leftrightarrow \hat{J}_{2}$); the maximally mixed state ($\frac{1}{2}\mathds{1}\leftrightarrow \hat{\mu}_{\frac{\mathds{1}}{2}}$); and the positive (negative) semi-definite matrix constraints with the non-negativity (negativity) element-wise restriction ($\geq(\leq) \leftrightarrow \underset{\mathrm{e.w.}}{\geq}(\underset{\mathrm{e.w.}}{\leq})$).

\end{widetext}

\end{document}